\documentclass[twocolumn,showpacs,preprintnumbers,amsmath,amssymb]{revtex4}
\usepackage{graphicx}% Include figure files
\usepackage{dcolumn}% Align table columns on decimal point
\usepackage{bm}% bold math

\def\be{\begin{equation}}
\def\ee{\end{equation}}
\def\bea{\begin{eqnarray}}
\def\eea{\end{eqnarray}}

\begin{document}

%\preprint{APS/123-QED}

\title{Adding to the paradox: the accelerated twin is older}

\author{Marek A. Abramowicz}%%%%%%%%%%%%%%%%%%%%%%%%%%%%%%%
\email{ marek.abramowicz@physics.gu.se}
\author{Stanis{\l}aw Bajtlik}%%%%%%%%%%%%%%%%%%%%%%%%%%%%%%
 \email{ bajtlik@camk.edu.pl}
\affiliation{Physics Department, G{\"o}teborg University,
SE-412-96 G{\"o}teborg, Sweden \\and N. Copernicus Astronomical
Center, Bartycka 18, PL-00-716 Warszawa, Poland}

\date{\today}

\begin{abstract}
We discuss a rather surprising version of the twin paradox in
which (contrary to the familiar classical version) the twin who
accelerates is {\it older} on the reunion than his never
acclerating brother.
\end{abstract}

\pacs{04.20.Cv}% PACS, the Physics and Astronomy
\maketitle

It is often claim that the resolution of the classical twin
paradox should be the acceleration of the ``traveling'' twin: he
must accelerate in order to turn around and meet his never
accelerating brother. The twin who accelerates is {\it younger} at
the reunion. Here we challenge this notion. We start with
describing a situation in which, like in the classical version of
the paradox, one of the twins accelerates, and the other one does
not accelerate. Quite contrary to what happens in the classical
version, the accelerated twin is {\it older} at the reunion. To
see that this is perfectly possible in rather familiar
circumstances, let us consider a static spacetime with the metric
(we use the $c = 1$ units),
%-----------------------------------------------------------------
\begin{equation}
\label{metric}
ds^2 = g_{tt}\,dt^2 +  g_{\phi\phi}\,d\phi^2 + g_{rr}\,dr^2 +
g_{\theta\theta}\,d\theta^2.
\end{equation}
%------------------------------------------------------------------
The metric tensor does not depend on time and azimuth, $\partial_t
g_{ik} = 0 = \partial_{\phi} g_{ik}$. It is $g_{tt} > 0$, while
all other $g_{ik}$ that appear in (\ref{metric}) are negative.
Although we keep our discussion general, and do not specify any
particular form of the metric $g_{ik}$, it could be convenient to
have in mind the Schwarzschild metric,
%-----------------------------------------------------------------
\begin{eqnarray}
\label{Schwarzschild}
&ds^2& = \left(1 - \frac{r_G}{r}\right)\,dt^2 \nonumber \\
&-& \left(1 - \frac{r_G}{r}\right)^{-1}\,dr^2 - r^2\,d\theta^2 -
r^2\sin^2\theta\,d\phi^2,
\end{eqnarray}
%------------------------------------------------------------------
where $r_G = 2GM$ is the gravitational radius and $M$ the mass of
the central gravitating body. The absolute standard of rest in the
Schwarzschild spacetime is given by the local Killing symmetry,
$\partial_t = 0$, the non-zero curvature, and the global condition
of the asymptotical flatness (the starry sky above). It is
convenient to define
%-----------------------------------------------------------------
\begin{equation}
\label{definitions}
\Phi = -\frac{1}{2}\ln (g_{tt}),~~R^2 =
-\frac{g_{\phi\phi}}{g_{tt}}.
\end{equation}
%------------------------------------------------------------------
On circular orbits the only non-zero components of the
four-velocity $u^i = dx^i/ds$ are $u^t$ and $u^{\phi} =
\Omega\,u^t$, with $\Omega = d\phi/dt$ being the orbital angular
velocity. The orbital velocity is $v = R\Omega$. Both $\Omega$ and
$v$ are measured with respect to the absolute standard of rest
mentioned above. On circular orbits one has,
%-----------------------------------------------------------------
\begin{equation}
\label{redshift-factor}
\frac{1}{u^t} =
{\rm e}^{-\Phi}{\sqrt{1 - v^2}}.
\end{equation}
%-----------------------------------------------------------------
The only non-zero radial component of the acceleration $a_k =
u^i\nabla_i u_k$ equals,
%-----------------------------------------------------------------
\begin{equation}
\label{acceleration-static-upper}
a_r \equiv a = \partial_r \Phi + \frac{v^2}{1 -
v^2}\left(\frac{\partial_r R}{R}\right).
\end{equation}
%-----------------------------------------------------------------
Note that one recovers the corresponding familiar Newton's formula
with $\Phi$ being the gravitational potential, $R$ being the
distance from the rotation axis, and with Newtonian velocity being
$V^2 = v^2/(1 - v^2)$. The acceleration is zero (geodesic motion)
when the orbital velocity takes its ``Keplerian'' value $v_K$
equal,
%-----------------------------------------------------------------
\begin{equation}
\label{acceleration-static-lower}
\frac{v^2_K}{1 - v^2_K} = - R\,\frac{\partial_r \Phi}{\partial_r
R}.
\end{equation}
%-----------------------------------------------------------------
Equation (\ref{acceleration-static-lower}) closely resembles its
Newtonian version.

Let us imagine two twins, A and B, located at a particular orbit
$r = r_0$, and orbiting with velocities $v_A$ and $v_B$. If $v_A
\not= v_B$, the twins will periodically meet. The ratio of their
proper times measured between two consecutive reunions is,
%-----------------------------------------------------------------
\begin{equation}
\label{ratio-proper-times}
\frac{\tau_A}{\tau_B} =
\frac{\int_A ds}{\int_B ds} =
\frac{\int_A (u^t)^{-1}dt}{\int_B (u^t)^{-1}dt} =
\frac{\sqrt{1 - v_A^2}}{ \sqrt{1 - v_B^2} }.
\end{equation}
%------------------------------------------------------------------
Suppose now that the twin A is not moving, $v_A = 0$, and the twin
B orbits with the Keplerian velocity, $v_B = v_K$. Thus, the twin
A is accelerated, $a_A = \partial_r \Phi \not= 0$, and the twin B
is not, $a_B = 0$. From (\ref{ratio-proper-times}) it follows that
$\tau_B = \tau_A {\sqrt{1 - v^2_K}} < \tau_A$ and therefore at the
reunion {\it the accelerated twin is older} than his
non-accelerated brother! Inspection of formula
(\ref{ratio-proper-times}) reveals that the ratio of proper times
measured by the twins between their two consecutive reunions
depends only on their orbital velocities $v_A$ and $v_B$, measured
with respect to the absolute global standard of rest.

Let us make an experiment in which the radius of the orbit is
always $r = r_0$, and also the absolute velocities of the two
twins remain unchanged, being always $v_A = 0$ and $v_B = v_K(r_0,
M_0) = v_0$, but the mass of the central body goes down from its
initial value $M= M_0$ to $M = 0$. Obviously, with the mass $M <
M_0$ and the orbital radius $r = r_0$, the velocity $v_0$ of the
twin B will not be Keplerian. Therefore, B (like his brother A)
will be accelerated. However, because the ratio of their proper
times depends not on accelerations, but on velocities, and these
do not change, the ratio will remain fixed. Thus, the moving twin
B will always be younger than his non-moving brother A,
%-----------------------------------------------------------------
\begin{equation}
\label{moving-is-younger}
\tau_B < \tau_A.
\end{equation}
%-----------------------------------------------------------------
The above relation is certainly true also in the strict limit $M =
0$, which corresponds to Minkowski spacetime. In this limit, we
recover the classical version the twin paradox: twin A is not
accelerated, and twin B is accelerated. It is the twin B who is
younger at reunions. But why is he younger? Should we say that in
the whole mass range $M \not =0$ the reason is his larger velocity
(independently of his acceleration) but only for $M =0$ the reason
changes and is his non-zero acceleration (independently of his
velocity, whatever velocity could mean)?
%-----------------------------------------------------------------
\begin{eqnarray}
\label{summarize-experiment}
&\lim_{M \rightarrow 0}\left(
\begin{tabular}{c}
{\rm the~higher~velocity~twin~is~younger}\\
{\rm acceleration~is~not~important}
\end{tabular}
    \right) = & \nonumber \\
&\left(
\begin{tabular}{c}
{\rm the~accelerated~twin~is~younger}\\
{\rm velocity~is~not~important}
\end{tabular}\right).&
\end{eqnarray}
%------------------------------------------------------------------
We agree with the opinion expressed previously  by several
scholars (the list of them starts notably with von Laue
\cite{von-laue-1913}) that contrary to the common perception, it
is not the acceleration which solves the twin paradox by breaking
symmetry between the two twins. A few scholars think that what
really breaks the symmetry here is the question who of the two
twins ``moves faster''. Indeed, in all non-classical versions of
the twin paradox considered so far, it was possible to define the
absolute standard of rest because the spacetime was equipped
either with a non-trivial topology or a global symmetry. The twin
who moved faster in this absolute sense, was younger.

{\it 1.~The twin paradox in compact spaces} was considered by
several authors, including Barrow and Levin \cite{barrow} (see
also e.g. \cite{bajtlik}). One may imagine a cylindrical Minkowski
spacetime, with a flat metric ${\rm diag}(1, -1, -1, -1)$,
compactified along one spatial direction. Inertial observers
moving in this direction along the same straight line meet
periodically. They may send light signals forwards and backwards
with respect to their motion, and measure the {\it
around-the-space} flight times $\Delta\,t_{FO}$, $\Delta\,t_{BC}$.
The observer for whom $\Delta\,t_{FO} = \Delta\,t_{BC}$ defines
the standard of the absolute rest. The faster another observer
moves with respect to this standard, the younger is he at the
reunion.

{\it 2.~The twin paradox on the photon sphere} was considered by
Abramowicz, Bajtlik, and Klu\'zniak \cite{photon-sphere}. Note
(after \cite{lasota, carter}), that from
(\ref{acceleration-static-upper}) it follows that the (non-zero)
acceleration is the same for all observers moving with an
arbitrary $v$ around the orbit characterized by the condition
$\partial_r R = 0$. One may check that this condition describes
the photon sphere, i.e. all circular null geodesics (orbits of
free photons). In the Schwarzschild spacetime the photon sphere is
given by $r = 3GM$, with $M$ being the mass of the gravity center.
Situation here is similar to that in the compact spaces case
(except that now all observers have the same, but non-zero,
acceleration): the younger at the reunion is the one who moves
faster.

{\it 3.~The twin paradox on intersecting orbits around a central
gravitating body} was considered by Holstein and Swift
\cite{geodesics}. Twins traveling on two different, but
intersecting, orbits would meet at least twice. They both have
zero acceleration. It was proved that the younger at the reunion
will be the one who traveled the longer path in space, i.e. the
one who (on average) was moving faster. Note that this prediction
of Einstein's relativity is verified experimentally. From
countless, and direct, measurements routinely done by the GPS
system (see e.g. \cite{gps}) we {\it know} that the twin orbiting
faster is younger.

To these three previously known examples, we add a new one, {\it
the twin paradox on circular orbits in a stationary, axially
symmetric spacetime.} Such spacetime has the metric,
%-----------------------------------------------------------------
\begin{equation}
\label{metric-non-static}
ds^2 = g_{tt}\,dt^2 + g_{\phi\phi}\,d\phi^2 + g_{t\phi}\,dt\,d\phi
+ g_{rr}\,dr^2 + g_{\theta\theta}\,d\theta^2,
\end{equation}
%------------------------------------------------------------------
with $\partial_t g_{ik} = 0 = \partial_{\phi} g_{ik}$. The
circular orbit and the angular velocity $\Omega$ are defined the
same way as previously for the static spacetime. The potential
$\Phi$, radius $R$, and velocity $v$ are now,
%-----------------------------------------------------------------
\begin{eqnarray}
\label{definitions-non-static}
\Phi &=& -\frac{1}{2}\ln \left( g_{tt} -
\omega^2\,g_{\phi\phi}\right)\\
R^2 &=& \frac{g_{\phi\phi}^2}{g_{t\phi}^2 - g_{tt}g_{\phi\phi}}\\
v &=& R\left(\Omega - \omega\right).
\end{eqnarray}
%-----------------------------------------------------------------
Here $\omega = -g_{t\phi}/g_{\pi\phi}$ is the angular velocity of
the dragging of inertial frames (the Lense-Thirring effect). From
these formulae one deduces that the ratio of proper times for the
two orbiting twins is, like in the static case,
%-----------------------------------------------------------------
\begin{equation}
\label{ratio-proper-times-non-static}
\frac{\tau_A}{\tau_B} =
\frac{\sqrt{1 - v_A^2}}{\sqrt{1 - v^2_B}}.
\end{equation}
%------------------------------------------------------------------
Thus, again, the faster twin is the younger. However, the
velocities of the twins are measured not with respect to the frame
of stationary observers connected to the Killing symmetry
$\partial_t = 0$, and equivalent to the standard of rest given by
the starry sky. In the non-static stationary spacetimes, the
standard of rest is where the ZAMO stays, i.e. the zero angular
momentum observers \cite{bardeen} who are rotating with respect to
the starry sky with the angular velocity $\omega$.

In conclusions, we stress that in all situations in which the
absolute motion may be defined in terms of some invariant {\it
global} properties of the spacetime, the twin who moves faster
with respect to the global standard of rest is younger at the
reunion, {\it irrespectively to twins' accelerations}. Could the
notion {\it ``the twin who moves faster is younger at the
reunion''} be somehow extended to the classical version of the
paradox in the Minkowski spacetime, for example by referring to
the starry sky above the twins? This question we should leave
unanswered.

\begin{acknowledgments}
We acknowledge the support from Polish State Committee for
Scientific Research, through grants NN203 394234 and
N203~009~31/1466.
\end{acknowledgments}

%\bibliography{twin}

\end{document}